\documentclass[conference]{IEEEtran}
\IEEEoverridecommandlockouts
\usepackage{cite}
\usepackage{amsmath,amssymb,amsfonts}
\usepackage{algorithmic}
\usepackage{graphicx}
\usepackage{textcomp}
\usepackage{xcolor}
\usepackage{algorithm}
\usepackage{algorithmic}
\usepackage{multirow}
\usepackage{subfigure}
\usepackage{setspace}
\usepackage{amsthm}
\def\BibTeX{{\rm B\kern-.05em{\sc i\kern-.025em b}\kern-.08em
    T\kern-.1667em\lower.7ex\hbox{E}\kern-.125emX}}
\begin{document}

\title{Diversifying Seeds and Audience in Social Influence Maximization
}

\author{\IEEEauthorblockN{Yu Zhang}
\IEEEauthorblockA{University of Illinois at Urbana-Champaign, Urbana, IL, USA \\
yuz9@illinois.edu}
}

\maketitle

\begin{spacing}{1}
\begin{abstract}
Influence maximization (IM) has been extensively studied for better viral marketing. However, previous works put less emphasis on how balancedly the audience are affected across different communities and how diversely the seed nodes are selected. In this paper, we incorporate audience diversity and seed diversity into the IM task. From the model perspective, in order to characterize both influence spread and diversity in our objective function, we adopt three commonly used utilities in economics (i.e., Perfect Substitutes, Perfect Complements and Cobb-Douglas). We validate our choices of these three functions by showing their nice properties. From the algorithmic perspective, we present various approximation strategies to maximize the utilities. In audience diversification, we propose a solution-dependent approximation algorithm to circumvent the hardness results. In seed diversification, we prove a ($1/e-\epsilon$) approximation ratio based on non-monotonic submodular maximization. Experimental results show that our framework outperforms other natural heuristics both in utility maximization and result diversification. 
\end{abstract}


\newcommand\simm{\textrm{Sim}}
\newcommand\dist{\textrm{Dis}}

\section{Introduction}
Viral marketing through social networks is becoming more and more popular due to the rapid growth of social media. Many advertisers promote their products by paying influential users, hoping the advertisement can propagate through reposts or shares. Kempe et al. \cite{kempe2003maximizing} first formulate this task as Influence Maximization (IM), in which they aim to obtain the largest influence spread with at most $k$ seed nodes.

Although IM and many of its variants have been extensively studied, most approaches disregard another concern -- \textit{diversity of nodes}. Diversity has been regarded as a crucial factor in many other tasks (e.g., document retrieval \cite{carbonell1998use} and node ranking \cite{he2012gender}). These tasks mainly focus on the diversity of selected items. In contrast, in IM, we argue that both the diversity of \textit{selected nodes} (i.e., seeds) and the diversity of \textit{activated nodes} (i.e., audience) need to be explored. 

\vspace{1mm}

\noindent \textbf{Audience Diversification.} In practical viral marketing scenarios, having diverse audience could bring many benefits. As we all know, a network contains multiple communities, and people in different communities may have different probabilities to buy the product after they see the advertisement. As mentioned in \cite{tang2014diversified}, if all targeted users come from one group or share the same feature, we are actually ``putting all eggs in one basket'', which will definitely increase the risk of marketing campaigns. 

\vspace{1mm}

\noindent \textbf{Seed Diversification.} Inspired by the widely observed \textit{homophily phenomenon} \cite{mcpherson2001birds,powell2005network}, Tang et al. \cite{tang2014diversified} relax the problem from enforcing
diversity on influenced crowd to just enforcing diversity on seed set. The idea behind this relaxation is that nodes with similar features are more likely to connect with each other. Therefore, if the seed set is diverse, the activated audience would be diverse as well. Besides, some real applications directly require the diversity of seed nodes. Consider the task of setting up a conference programming committee \cite{mei2010divrank}. The goal is to invite influential researchers from all related areas. Without diversity, the interest of the committee could be biased.

In this paper, we study \textit{Audience-Diversified IM} (ADIM) and \textit{Seed-Diversified IM} (SDIM) in a unified framework. In ADIM, we model influence spread in each community as a factor in the objective function. Intuitively, the objective should satisfy two properties: (1) \textit{Pareto Efficiency}: Activating more nodes in one community while keeping the spreading results in other communities will increase the objective value, and (2) \textit{Community-Level Balance}: Fixing the total influence spread, a more equally distributed spreading result over the communities will have a higher objective value. In SDIM, we model overall spread and seed diversity as two factors in our objective. Again, the selection of objectives should follow two principles: (1) \textit{Pareto Efficiency}: Increasing either spread or diversity will boost the objective value, and (2) \textit{Spread-Diversity Tradeoff}: When diversity is lower, we are more willing to substitute spread for diversity.

Inspired by economics practice \cite{varian2014intermediate}, we propose to use three kinds of utility functions, \textit{Perfect Substitutes}, \textit{Perfect Complements} and \textit{Cobb-Douglas} to composite multiple factors into the final objective. We first prove that the three utility functions all satisfy the properties mentioned above. Then we propose algorithms to maximize these objectives case by case. In ADIM, we prove that for Perfect Substitutes, the hill-climbing greedy method can provide a $(1-1/e-\epsilon)$ approximation guarantee. For Perfect Complements and Cobb-Douglas, we show that ADIM is hard to approximate with any positive constant ratio. Given the hardness, we propose an algorithm with solution-dependent approximation guarantees using the \textit{Sandwich Approximation} strategy \cite{lu2015competition,chen2016robust,lin2017boosting}. 
In SDIM, we prove the three utilities are all submodular (but not monotonic). Therefore, the \textit{Random Greedy} strategy \cite{buchbinder2014submodular} for size-constrained non-monotonic submodular optimization can guarantee a $(1/e-\epsilon)$ approximation ratio.


We conduct extensive experiments on two real-world networks with different choices of utility functions. Our method consistently outperforms several baselines including traditional IM \cite{kempe2003maximizing}, diversified node ranking \cite{he2012gender} and diversified IM \cite{tang2014diversified} in both ADIM and SDIM. Besides, putting the utilities aside, we prove the success of our diversified IM framework from the views of \textit{entropy} \cite{tang2014diversified} and \textit{coverage} \cite{zhu2007improving}, indicating that the result is truly diversified, not just maximizing an objective. 

\section{Related Work}
\noindent \textbf{Influence Maximization.} Kempe et al. \cite{kempe2003maximizing} first formalize IM as a discrete optimization problem. Subsequent efforts following their framework can be divided into two directions. In one direction, researchers focus on how to accelerate the vanilla hill-climbing greedy algorithm \cite{leskovec2007cost,chen2009efficient,chen2010scalable,chen2010scalable1,borgs2014maximizing,tang2014influence,tang2015influence,tang2018online}. In the other direction, researchers propose new problem settings \cite{liu2012time,chen2015online,chen2016robust,he2016robust,zhang2017top} based on IM. Due to space limitation, we do not list all the variants here. One can refer to a recent tutorial \cite{aslay2018influence} for more details. Despite their success, the final cascade size is their primary criterion in selecting influential nodes. In contrast, our model incorporates diversity as a factor.

\vspace{1mm}

\noindent \textbf{Seed Diversification.} Diversifying ranked items is first studied in information retrieval \cite{carbonell1998use}. Subsequent efforts \cite{agrawal2009diversifying,gollapudi2009axiomatic,borodin2012max,ceccarello2018fast} propose different models to describe the tradeoff between relevance and diversity. Some of these methods show their power in IR, but they hinge on specific choices of similarity functions and cannot be easily generalized to social network scenarios. \cite{zhu2007improving}, \cite{mei2010divrank} and \cite{tong2011diversified} then transfer this framework to social networks for diversified node ranking. He et al. \cite{he2012gender} further generalize the framework to a wider range of relevance and similarity functions. However, they require the authority part to be modular (i.e., authority$(\{u,v\})$ is the sum of authority$(\{u\})$ and authority$(\{v\})$), which does not hold for IM. Besides, some studies mentioned above require the dissimilarity function between two elements to form a metric, which is not true in some settings (e.g., the community and embedding settings discussed below).

\vspace{1mm}

\noindent \textbf{Audience Diversification.}
\cite{garimella2017balancing,aslay2018maximizing,matakos2018tell} study how to maximize the diversity of \textit{exposure} in a network. Their goal is to recommend diverse content to the audience so that friends can have different knowledge. This is different from our goal to diversify the influenced nodes with the same piece of content. To the best of our knowledge, \cite{tang2014diversified} is the only previous work exploring audience/seed diversity in social influence maximization. However, their objective is essentially a weighted sum of spread and diversity. In contrast, our framework systematically studies a family of utility functions from the perspective of economics \cite{varian2014intermediate}.

\section{Problem Formulation}
\subsection{Audience-Diversified Influence Maximization (ADIM)}
\noindent \textbf{Influence Maximization (IM) \cite{kempe2003maximizing}.} In a network $G=(V,E)$, an information cascade starts from an initially active set $S \subseteq V$. Then the information propagates in $G$ under certain spreading models as the time runs forward. There are two spreading models \cite{kempe2003maximizing} commonly studied in IM.

In the \textit{Independent Cascade (IC)} model, when a node $u$ becomes active at time $t$, it gets a chance to activate each of its inactive neighbor $v$ at time $t+1$, with probability $p_{uv}$. If $v$ is not influenced by $u$, $u$ cannot make any further attempts in the subsequent rounds.

In the \textit{Linear Threshold (LT)} model, each node $v$ has a threshold $\theta_v \sim U[0,1]$ and each edge $(u,v)$ has a weight $b_{uv}$. If at time $t$, we have $\sum_{v\textrm{'s active neighbor }u}b_{uv}\geq \theta_v$ for an inactive node $v$, then $v$ will become active at time $t+1$.

Given the spreading model, the expected influence spread is defines as
\begin{equation}
\sigma(S) = \mathbb{E}[|S_{\tt act}|], \notag
\end{equation}
where $S_{\tt act}$ is the set of nodes activated in the cascade (including $S$). IM aims to find an $S$ with at most $k$ nodes to maximize $\sigma(S)$.

\vspace{1mm}

\noindent \textbf{Audience Diversity.} To introduce audience diversity, we assume there are $C$ communities (i.e., audience groups) $V_1, V_2, ..., V_C$ in a network, where $V_c$ is the set of nodes in the $c$-th community. Similar to IM, the expected influence spread in $V_c$ is
\begin{equation}
    \sigma_c(S) = \mathbb{E}[|S_{\tt act} \cap V_c|]. \notag
\end{equation}
Considering influence spread in each community as a factor, the objective function can be defined as
\begin{equation}
    f(S) = F\Big(\alpha_1\sigma_1(S),\alpha_2\sigma_2(S), ..., \alpha_C\sigma_C(S)\Big), \notag
\end{equation}
where $\alpha_c$ is the weight of community $V_c$ in the utility. For example, if we would like to study the ``proportion'' of activated nodes in each community instead of the absolute number, we can set $\alpha_c = 1/|V_c|$.

Intuitively, $F$ should satisfy the following two properties:

\vspace{1mm}

\noindent \textbf{(P1) Pareto Efficiency.} $\forall c$ and $\epsilon \geq 0$, $F(x_1,...,x_{c-1},x_c$ $,x_{c+1},...,x_C) \leq F(x_1,...,x_{c-1},x_c+\epsilon,x_{c+1},...,x_C)$.

(P1) models the fact that we favor an action which benefits at least one community without making the others worse off.

\vspace{1mm}

\noindent \textbf{(P2) Community-Level Balance.} $\forall k$ $(1 \leq k \leq C)$ communities (w.l.o.g., $V_1,...,V_k$), $F(x_1,...,x_k,x_{k+1},...x_C) \leq F(\frac{x_1+...+x_k}{k},...,\frac{x_1+...+x_k}{k},x_{k+1},...,x_C)$.

(P2) models our preference for a more equally distributed (i.e., diversified) spreading result over communities.

\vspace{1mm}

Mathematically, various candidates satisfy (P1) and (P2). In this paper, inspired by the utility functions in economics, we focus on three cases which are commonly adopted to characterize users' preference when there are multiple factors \cite{varian2014intermediate}.

\vspace{1mm}

\noindent \textbf{Perfect Substitutes (Linear Utility).} Two goods are perfect substitutes if the user is willing to substitute one good for the other at a constant rate (e.g., for most people, Pepsi and Coke). In mathematics, the function is essentially a weighted sum.
\begin{equation}
 f_S(S) = \sum_{c=1}^C \alpha_c \sigma_c(S), \ \ \ \ (\alpha_c > 0). \notag
\end{equation}

Since $f_S(\cdot)$ is very similar to the global spread $\sigma(\cdot)$, we consider a more generalized form called \textit{Constant Elasticity of Substitution} (CES) \cite{varian2014intermediate}.
\begin{equation}
 f_{CES}(S) = \Big(\sum_{c=1}^C (\alpha_c \sigma_c(S))^\rho\Big)^{1/\rho}, \ \ (0 < \rho \leq 1,\  \alpha_c > 0). \notag
\end{equation}

\noindent \textbf{Perfect Complements (Leontief Utility).} A nice example is that of left shoes and right shoes. if we have exactly two pairs of shoes, then neither extra left shoes nor extra right shoes can do us a bit of good. When the influence spread in different communities are regarded as perfect complements, we have 
\begin{equation}
 f_C(S) = \min_{1\leq c\leq C} \alpha_c\sigma_c(S), \ \ \ \ (\alpha_c > 0). \notag
\end{equation}

\noindent \textbf{Cobb-Douglas Utility.} In economics, the utility function usually follows the law of diminishing returns: Adding more of one factor, while holding all other constant, will yield lower incremental per-unit returns. Cobb-Douglas utility is commonly used to describe
this property: 
\begin{equation}
 f_D(S) = \prod_{c=1}^C (\alpha_c\sigma_c(S))^{1/C}, \ \ \ \ (\alpha_c > 0). \notag
\end{equation}

\vspace{1mm}

The three kinds of utilities correspond to $F_S(x) = \sum_c x_c$ (more generally, $F_{CES}(x) = (\sum_c x_c^\rho)^{1/\rho}$), $F_C(x) = \min_c x_c$ and $F_D(x) = (\prod_c x_c)^{1/C}$, respectively. 

\vspace{1mm}

\noindent \textbf{Theorem 1.} \textit{$F_{CES}\ (0< \rho \leq 1)$, $F_C$ and $F_D$ all satisfy (P1) and (P2).}

\vspace{1mm}

\noindent \textbf{Definition 1 (ADIM).} 
\textit{In a network $G=(V,E)$, given a size constraint $k$ and a utility function $f \in \{f_{CES}\ (0 < \rho \leq 1), f_C, f_D\}$,
$\max_{|S| = k} f(S)$. }

\vspace{1mm}

\subsection{Seed-Diversified Influence Maximization (SDIM)}
\label{sec:sdim}
Audience diversity can be \textit{implicitly} described by the spreading results in all communities. In contrast, it is difficult to characterize seed diversity reversely using influence spread. Therefore, we present an \textit{explicit} definition of seed diversity.

\vspace{1mm}

\noindent \textbf{Seed Diversity.}
Assume we have a pairwise node similarity function $\simm(\cdot, \cdot) \in [0,1]$. Inspired by the studies in diversified ranking \cite{gollapudi2009axiomatic,borodin2012max,he2012gender}, we define seed diversity as the average pairwise dissimilarity in $S$.
\begin{equation}
d(S) = \frac{1}{|S|(|S|-1)}\sum_{u,v \in S, u\neq v}\Big(1-\simm(u,v)\Big).  \notag
\end{equation}

The form of $\simm(\cdot, \cdot)$ can be specified from various perspectives (e.g., semantics, types, etc.). Following the setting in ADIM, we partition the network into $C$ communities. If the communities are \textit{overlapping}, BigCLAM \cite{yang2013overlapping} uses a vector $F_u = [F_{u1}, ..., F_{uC}]^T$ to represent node $u$, where $F_{uc}$ is the probability that $u$ belongs to community $V_c$. In \cite{yang2013overlapping}, the proximity between $u$ and $v$ is defined as 
\begin{equation}
\simm_C(u,v) = 1-\exp(-F_u^TF_v). \notag
\end{equation}
The same formula can be adopted for \textit{disjoint} communities (in which $F_u$ becomes a ``one-hot'' vector), and we will have $\simm_C(u,v) = 1-1/e$ if $u$ and $v$ belong to the same community, and 0 otherwise.


We can also follow the popular \textit{node embedding} setting to define $\simm(\cdot, \cdot)$. For each node $u$, a low-dimensional vector $e_u$ is learned to preserve the proximity in the original network. There are several well-known node embedding algorithms \cite{perozzi2014deepwalk,tang2015line,grover2016node2vec}, among which LINE \cite{tang2015line} explicitly defines the proximity between two nodes. In (first-order) LINE, the proximity is defined as 
\begin{equation}
\simm_L(u,v) = \frac{1}{1+\exp(-e_u^Te_v)}. \notag
\end{equation}

We would like to mention that our dissimilarity function $1-\simm(\cdot,\cdot)$ \textit{need not} be a metric. For example, it is easy to check that neither $1-\simm_C(u,u)$ nor $1-\simm_L(u,u)$ is 0.

\vspace{1mm}

Following ADIM, we define an objective function that jointly models influence spread and seed diversity as two factors.
\begin{equation}
    g(S) = G\Big(\sigma(S),\ \beta\cdot d(S)\Big), \notag
\end{equation}
where $\beta$ is a constant factor making $\beta\cdot d(S)$ share the same magnitude with $\sigma(S)$ (e.g., $\beta=|V|$).

Intuitively, $G$ should also satisfy (P1) with two variables. Besides, we propose the following property to characterize our willingness to substitute spread for diversity.

\vspace{1mm}

\noindent \textbf{(P3) Spread-Diversity Tradeoff.} If $\exists\ \epsilon, \delta \geq 0$ such that $G(x_1-\epsilon, x_2+\delta) = G(x_1, x_2)$, then $G(x_1-2\epsilon, x_2+2\delta) \leq G(x_1-\epsilon, x_2+\delta)$.

(P3) tells us that when diversity is lower (i.e., $x_2$), we are more willing to substitute spread for diversity (i.e., at the rate of $\epsilon/\delta$). When diversity becomes higher (i.e., $x_2+\delta$), we no longer expect the substitution at the same rate. In economics, this is named as the law of diminishing marginal rates of substitution \cite{varian2014intermediate}.

We can still adopt the three utility functions used in ADIM.

\vspace{1mm}

\noindent \textbf{Perfect Substitutes.} When $G_S(x_1, x_2) = x_1+x_2$, we have
\begin{equation}
g_S(S) = \sigma(S) + \beta\cdot d(S), \ \ \ \ (\beta > 0). \notag
\end{equation}
Essentially, Perfect Substitutes is a weighted sum of spread and diversity. \cite{gollapudi2009axiomatic} and \cite{borodin2012max} also studied this utility function for diversified ranking. However, in their models, $1-\simm(\cdot,\cdot)$ must form a metric. Without this assumption (e.g., in our community and embedding settings), their algorithms do not have approximation guarantees. 

\vspace{1mm}

\noindent \textbf{Perfect Complements.} When $G_C(x_1, x_2) = \min\{x_1,x_2\}$,
\begin{equation}
g_C(S) = \min\{\sigma(S),\ \beta\cdot d(S)\}, \ \ \ \ (\beta > 0). \notag
\end{equation}

\vspace{1mm}

\noindent \textbf{Cobb-Douglas Utility.} When $G_D(x_1, x_2) = x_1^ax_2^b$ $(0 < a,b \leq 1$ and $a+b = 1)$, we have
\begin{equation}
g_D(S) = \sigma(S)^a\cdot (\beta d(S))^b \propto \sigma(S)^a \cdot d(S)^b. \notag
\end{equation}

\vspace{1mm}

\noindent \textbf{Theorem 2.} \textit{$G_S$, $G_C$ and $G_D$ all satisfy (P1) and (P3).}

\vspace{1mm}

Note that (P2) and (P3) are not equivalent. For example, $x_1^{0.9}x_2^{0.1}$ satisfies (P3) but violates (P2).

\vspace{1mm}

\noindent \textbf{Definition 2 (SDIM).} 
\textit{In a network $G=(V,E)$, given a size constraint $k$ and a utility function $g \in \{g_S, g_C, g_D\}$,
$\max_{|S| = k} g(S)$. }

\vspace{1mm}

It is easy to show that both ADIM and SDIM can be viewed as the extension of traditional IM, so they are \texttt{NP}-hard.

\section{Algorithms}

Due to \texttt{NP}-hardness, we focus on finding approximation algorithms for ADIM and SDIM. Table \ref{tab:thm} summarizes our results in this section.

\begin{table}[]
\centering
\caption{Approximation results for ADIM and SDIM.}
\vspace{-0.5em}
\scalebox{0.85}{
\begin{tabular}{c|c|c}
\hline
Utility      & Audience Diversification                                                       & Seed Diversification                                           \\ \hline
Substitutes  & \begin{tabular}[c]{@{}c@{}}$1-1/e-\epsilon$\\ $(1-1/e-\epsilon)^{1/\rho}$ for CES\end{tabular}                      & $1/e-\epsilon$ \\ \hline
Complements  & \begin{tabular}[c]{@{}c@{}}\texttt{NP}-Hard to Approx.\\ Solution-dependent Approx.\end{tabular}  & $1/e-\epsilon$ \\ \hline
Cobb-Douglas & \begin{tabular}[c]{@{}c@{}}\texttt{NP}-Hard to Approx.\\ Solution-dependent Approx.\end{tabular} & $1/e-\epsilon$ \\ \hline
\end{tabular}
}
\vspace{0em}
\label{tab:thm}
\end{table}

\subsection{Audience-Diversified Influence Maximization}
\noindent \textbf{Perfect Substitutes and CES.}
With the help of submodularity, Perfect Substitutes and CES $(0 < \rho \leq 1)$ can be solved via the traditional hill-climbing greedy method.

\vspace{1mm}

\noindent \textbf{Lemma 1 \cite{tang2014diversified}.} 
\textit{Under IC or LT model, $\sigma_c(S)$ is monotonic and submodular for any $c=1,...,C$.}

\vspace{1mm}

\noindent \textbf{Theorem 3.}
\textit{Under IC or LT model, {\sc Greedy}($f_{CES}^\rho,k$) achieves a $(1-1/e-\epsilon)^{1/\rho}$ approximation guarantee when $0< \rho \leq 1$.}

\vspace{1mm}



When $\rho = 1$, we have the common $(1-1/e-\epsilon)$ approximation ratio for Perfect Substitutes.

\vspace{1mm}

\noindent \textbf{Perfect Complements.}
Algorithm 1 cannot be applied to $f_C(\cdot)$ and $f_D(\cdot)$ because neither of them is submodular. In fact, it is hard to obtain any positive constant approximation guarantee in these two cases.

\vspace{1mm}

\noindent \textbf{Theorem 4.}
\textit{Under IC model, ADIM is {\tt NP}-hard to approximate with any positive constant factor for $f_C(\cdot)$ and $f_D(\cdot)$. }

\vspace{1mm}

To circumvent this hardness result, we attempt to prove a \textit{solution-dependent} guarantee \cite{lu2015competition,lin2017boosting}. In light of the \textit{Sandwich Approximation} (SA) strategy \cite{lu2015competition}, we propose Algorithm \ref{alg:sag} that works for both $f_C(\cdot)$ and $f_D(\cdot)$.

SA aims to optimize a submodular upper bound of the original objective (or a lower bound \cite{lin2017boosting}, or both \cite{lu2015competition}). To be specific, we look for a submodular function $f_C^+$, where $f_{C}(S)$ is always smaller than $f_C^+(S)$. In the case of Perfect Complements, there are $C$ upper bounds $\alpha_i\sigma_i(S)$ $(i = 1,2,...,C)$, each of which is monotonic and submodular. When we apply the SA strategy on all of these $C$ upper bounds, the following result can be derived.

\vspace{1mm}

\noindent \textbf{Theorem 5.}
\textit{Under IC or LT model, {\sc Upper-Greedy}$(f_C,$ $\{\alpha_1\sigma_1,$ $...,\alpha_C\sigma_C\}, k)$ finds a seed set $S$ and guarantees that 
\begin{equation}
f_C(S) \geq \max_{1\leq i \leq C}\frac{f_C(S_i)}{\alpha_i\sigma_i(S_i)}(1-\frac{1}{e}-\epsilon)f_C(S_C^*), \notag
\end{equation}
where $S_C^*$ is the optimal solution for $f_C(\cdot)$.}

\vspace{1mm}



$\max_{1\leq i \leq C}\frac{f_C(S_i)}{\alpha_i\sigma_i(S_i)}(1-\frac{1}{e}-\epsilon)$ is referred as a solution-dependent approximation ratio since it is related to $S_i$. Note that it can be calculated once we have the solution, and the true effectiveness of \textsc{Upper-Greedy} depends on the gap between $\alpha_i\sigma_i$ and $f_C$. In our case, when there is only one community, $f_C \equiv \alpha_1\sigma_1$. Then Algorithm \ref{alg:sag} has the common $(1-1/e-\epsilon)$ approximation ratio.

\vspace{1mm}

\noindent \textbf{Cobb-Douglas.} Again, we adopt the SA strategy. Our selection of the upper bound is $f_D^+(S) = \frac{1}{C}\sum_{c=1}^C \alpha_c \sigma_c(S)$. Since the geometric mean is always less than or equal to the arithmetic mean, we have $f_D^+(S)\geq f_D(S)$. Similar to Theorem 5, the following result holds.

\begin{algorithm}[!t]
\small
\caption{\textsc{Greedy}($f, k$)}
\label{alg:greedy}
\begin{algorithmic}[1]
\STATE initialize $S = \emptyset$
\FOR {$i = 1$ to $k$}
     \STATE select $u = \arg\max_{v\in V\backslash S}(f(S\cup\{v\})-f(S))$
     \STATE $S = S\cup\{u\}$
\ENDFOR
\STATE output $S$
\end{algorithmic}
\end{algorithm}

\begin{algorithm}[t]
\small
\caption{\textsc{Upper-Greedy}($f, \{f^+_1,...,f^+_U\}, k$)}
\label{alg:sag}
\begin{algorithmic}[1]
\STATE $S_0 =\ $\textsc{Greedy}$(f, k)$
\FOR {$i = 1$ to $U$}
    \STATE $S_i =\ $\textsc{Greedy}$(f^+_i, k)$
\ENDFOR
\STATE $S = \arg\max_{0\leq i \leq U}f(S_i)$
\STATE output $S$
\end{algorithmic}
\end{algorithm}

\vspace{1mm}

\noindent \textbf{Theorem 6.}
\textit{Under IC or LT model, {\sc Upper-Greedy}$(f_D,$ $\{f_{D+}\},$ $k)$ finds a seed set $S$ and guarantees that 
\begin{equation}
f_D(S) \geq \frac{f_D(S_1)}{f_D^+(S_1)}(1-\frac{1}{e}-\epsilon) f_D(S_D^*). \notag
\end{equation}
where $S_D^*$ is the optimal solution for $f_D(\cdot)$.}

\vspace{1mm}


Again, when there is only one community, $f_D$ and $f_{D+}$ are equivalent, and we get the common $(1-1/e-\epsilon)$ approximation ratio.

\vspace{1mm}

\noindent \textbf{Time Complexity.} The time complexity of \textsc{Greedy} is $O(k|V|\mathcal{T})$, where $\mathcal{T}$ is the time to calculate $f(S)$, or $\sigma_c(S)\ (c=1,...,C)$. Chen et al. \cite{chen2010scalable,chen2010scalable1} have pointed out the hardness of this computation under IC and LT models, but an arbitrarily small $\epsilon$ can be obtained through Monte Carlo simulation. Suppose we run $M$ trials of simulation, since each iteration takes $O(|V|)$ time, the overall time complexity will be $O(kM|V|^2)$. Similarly, the time complexity of \textsc{Upper-Greedy} is $O(kUM|V|^2)$, where $U$ is the number of upper bounds used ($U=C$ for Perfect Complements and $U=1$ for Cobb-Douglas). 
In this paper, we do not focus on the efficiency of estimating influence spread. However, it is worth noting that Monte Carlo simulations can be replaced by Reverse Influence Sampling strategies \cite{borgs2014maximizing,tang2014influence,tang2015influence} to accelerate our algorithms.

\subsection{Seed-Diversified Influence Maximization}
Now we proceed to SDIM. Recall that $g_S(S)$, $g_C(S)$ and $g_D(S)$ are all composites of $\sigma(S)$ and $d(S)$. Although $\sigma(S)$ has good properties under IC and LT models, $d(S)$ can be neither monotonic nor submodular. To tackle this issue, we consider a problem equivalent to SDIM. 

We already know that
\begin{equation}
\begin{split}
d(S) &= \frac{1}{|S|(|S|-1)}\sum_{u,v \in S, u\neq v}\Big(1-\simm(u,v)\Big) \\
&= 1 -  \frac{1}{|S|(|S|-1)}\sum_{u,v \in S, u\neq v}\simm(u,v).\notag
\end{split}
\end{equation}
Now we consider
\begin{equation}
\tilde{d}(S) = 1 -  \frac{1}{k(k-1)}\sum_{u,v \in S, u\neq v}\simm(u,v).\notag
\end{equation}
Note that $d(S) = \tilde{d}(S)$ when $|S| = k$. Therefore, for any $\tilde{g}(S) = G(\sigma(S), \tilde{d}(S))$,
\begin{equation}
\max_{|S|=k} g(S) \ \ \Longleftrightarrow \ \ \max_{|S|=k} \tilde{g}(S).\notag
\end{equation}

Following this way, our SDIM problem becomes maximizing $\tilde{g}_S(S) = \sigma(S) + \beta\tilde{d}(S)$, $\tilde{g}_C(S) = \min\{\sigma(S),\beta\tilde{d}(S)\}$ and $\tilde{g}_D(S) = \sigma(S)^a\tilde{d}(S)^b$ subject to $|S| = k$.

Note that we study $\tilde{d}(S)$ instead of $d(S)$ because it has better properties.

\vspace{1mm}

\noindent \textbf{Lemma 2.} \textit{For any $\simm{(\cdot,\cdot)} \in [0,1]$, $\tilde{d}(S)$ is non-negative, decreasing and submodular.}

\vspace{1mm}

Consequently, we can prove the following.

\vspace{1mm}

\noindent \textbf{Theorem 7.}
\textit{For any monotonic and submodular $\sigma(\cdot)$ and any $\simm{(\cdot,\cdot)} \in [0,1]$, $\tilde{g}_S(S)$, $\tilde{g}_C(S)$ and $\tilde{g}_D(S)$ are all non-negative and submodular. }

\vspace{1mm}

Theorem 7 naturally applies to spreading functions under IC and LT models, and to diversity functions in our community detection and node embedding settings. With this Theorem, we successfully transform SDIM to a size-constrained non-monotonic submodular maximization problem, where we are able to adopt the \textsc{Random-Greedy} algorithm (Algorithm 3) proposed in \cite{buchbinder2014submodular}. \textsc{Random-Greedy} is a natural generalization of the vanilla greedy algorithm. Instead of picking the best single node in each iteration, it first finds $k$ nodes with the highest marginal gains and then randomly selects one node from the top-$k$ candidates to add. The following result is proved in \cite{buchbinder2014submodular}.

\vspace{1mm}

\noindent \textbf{Theorem 8 \cite{buchbinder2014submodular}.}
\textit{Let $g(\cdot)$ be a non-negative submodular (not necessarily monotonic) function. For the problem $\max_{|S| = k}g(S)$,}

\textit{(1) {\sc Random-Greedy}$(g, k)$ finds a set $S$ and guarantees $\mathbb{E}[g(S)]\geq \max\{0.266, \frac{1}{e}(1-\frac{k}{e|V|})\}\cdot g(S^*)$, where $S^*$ is the optimal solution.}

\textit{(2) There is another {\sc Continuous-Double-Greedy} algorithm. By taking the better of the outputs of {\sc Random-Greedy} and this algorithm, we can guarantee that $\mathbb{E}[g(S)]\geq 0.356\cdot g(S^*)$.}

\vspace{1mm}

When $k = o(|V|)$, the approximation rate of \textsc{Random-Greedy} becomes $\max\{0.266, 1/e-o(1)\} = 1/e-\epsilon > 0.356$. In this case, there is no need to take the ``better'' of the two
algorithms since it does not give us a better approximation guarantee. 
For influence maximization or influential node mining, we usually have $k=o(|V|)$. (In the real world, we can hardly obtain an initial seed set whose size is proportional to the whole network size, and when we talk about ``influential nodes'', we may not need $O(|V|)$ candidates.) Therefore, we just use \textsc{Random-Greedy} due to its simplicity and efficiency. We also assume $k \ll |V|$ in all of our experiments.

Putting Theorems 7 and 8 together, we get a $(1/e - \epsilon)$ approximation algorithm for SDIM. 

\vspace{1mm}

\noindent \textbf{Time Complexity.} The time complexity of \textsc{Random-Greedy} is $O(k|V|\mathcal{T})$, where $\mathcal{T}$ is the time to calculate $\tilde{g}(S)$, or $\sigma(S)$ and $\tilde{d}(S)$. Incrementally updating $\tilde{d}(S)$ only requires $O(k)$ time. For $\sigma(S)$, as mentioned above, it can be approximated through Monte Carlo simulation. Therefore, the overall complexity is $O(kM|V|^2)$. 
Again, Reverse Influence Sampling strategies can be applied to devise a more efficient version of \textsc{Random-Greedy}.

\begin{algorithm}[t]
\small
\caption{\textsc{Random-Greedy}($f$, $k$)}
\label{alg:Framework}
\begin{algorithmic}[1]
\STATE initialize $S_0 = \emptyset$
\FOR {$i = 1$ to $k$}
     \STATE Let $M_i \subseteq V-S_{i-1}$ be the subset of size $k$ maximizing $\sum_{v\in M_i}f(S_{i-1}\cup\{v\})-f(S_{i-1})$
     \STATE Randomly select $u$ from $M_i$
     \STATE $S_{i} = S_{i-1}\cup\{u\}$
\ENDFOR
\STATE output $S_k$
\end{algorithmic}
\end{algorithm}

\section{Experiments}

We aim to answer three questions in our experiments: (1) Can we achieve higher utilities in comparison with baseline algorithms? (2) Putting the utilities aside, in ADIM, can we really diversify the activated crowd without hurting the spread? (3) Similarly, in SDIM, can we diversify the selected seeds with little reduce in their influence power?

\subsection{Experimental Setup}

\noindent \textbf{Datasets.} Two benchmark networks are used: (1) \textsc{FourArea} \cite{sun2011pathsim} is an academic collaboration network extracted from DBLP. It contains authors from 4 areas: \textit{database}, \textit{data mining}, \textit{machine learning} and \textit{information retrieval}. (2) \textsc{Epinions} \cite{richardson2003trust} is a who-trust-whom network of a consumer review site Epinions.com. We adopt BigCLAM \cite{yang2013overlapping} to detect 10 overlapping communities in the network. Note that in both datasets, there are nodes not belonging to any community. We summarize the dataset statistics in Table \ref{tab:stat}.
\vspace{-0.8em}

\begin{table}[H]
\footnotesize
\centering
\caption{Dataset Statistics.}
\vspace{-0.5em}
\scalebox{0.9}{
\begin{tabular}{cccccc}
\hline
Dataset  & $|V|$    & $|E|$     & Edge Type  & $C$  & Community Type \\ \hline
\textsc{FourArea} \cite{sun2011pathsim} & 27,199 & 66,832  & Undirected & 4  & Disjoint       \\
\textsc{Epinions} \cite{richardson2003trust} & 75,879 & 508,837 & Directed   & 10 & Overlapping    \\ \hline
\end{tabular}
}
\label{tab:stat}
\end{table}

\begin{figure}[t]
\centering
  \subfigure[\textsc{FourArea}, CES]{
    \includegraphics[scale=0.43]{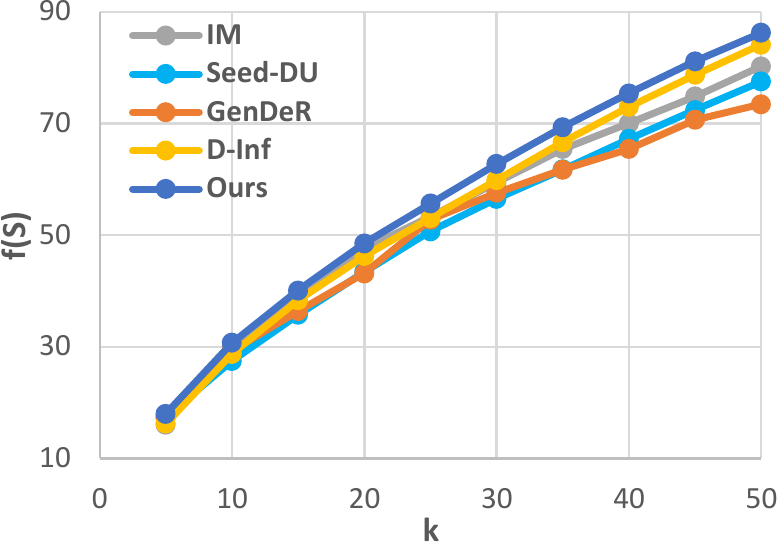}}
  \hspace{2ex}
  \subfigure[\textsc{Epinions}, CES]{
    \includegraphics[scale=0.43]{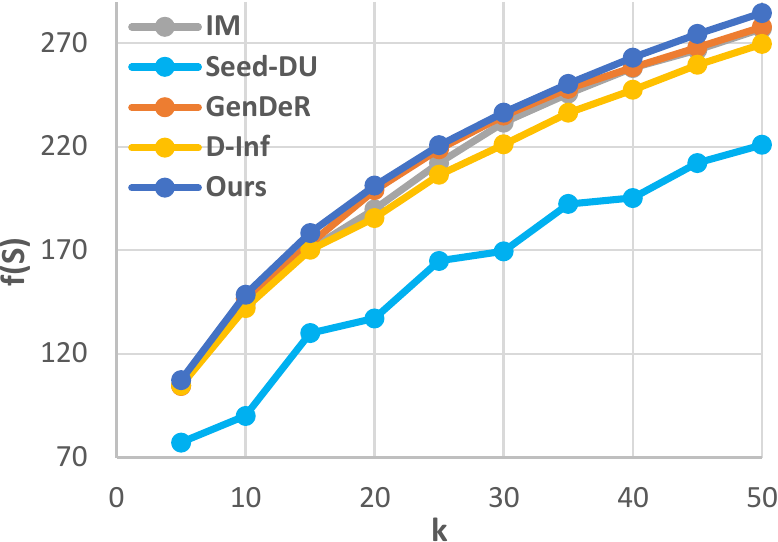}}
  \vspace{-2mm}
  \subfigure[\textsc{FourArea}, PC]{
    \includegraphics[scale=0.43]{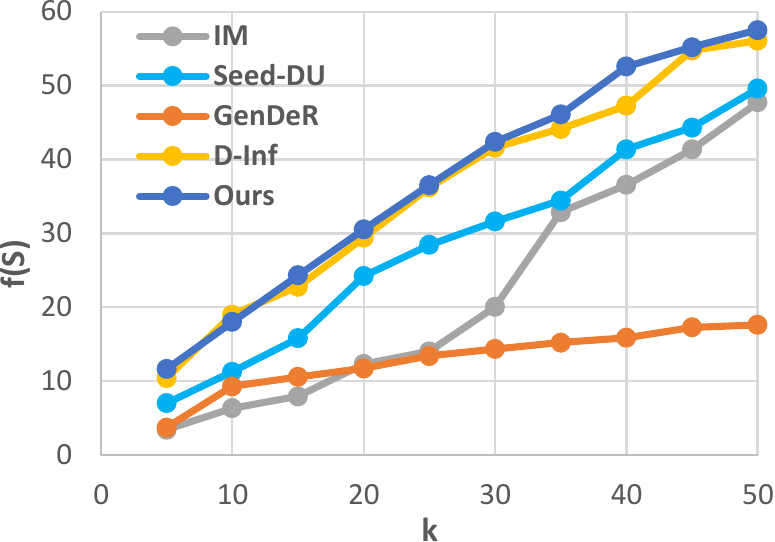}}
  \hspace{2ex}
  \subfigure[\textsc{Epinions}, PC]{
    \includegraphics[scale=0.43]{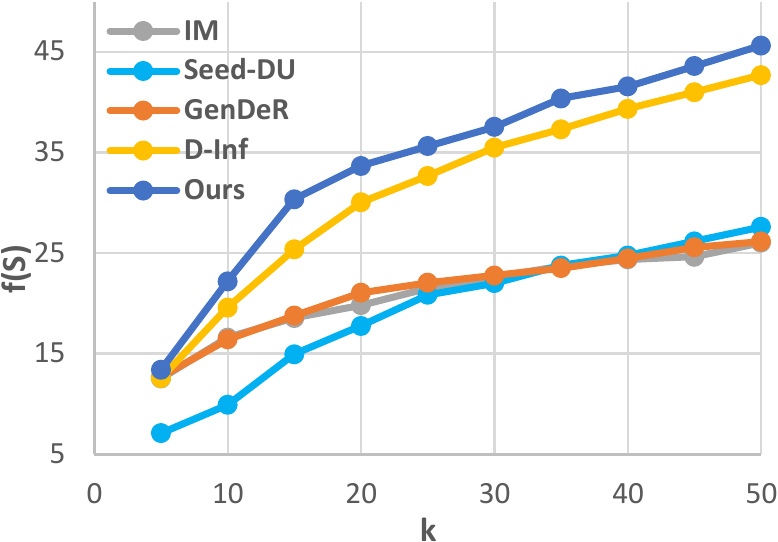}}
  \vspace{-2mm}
  \subfigure[\textsc{FourArea}, CD]{
    \includegraphics[scale=0.43]{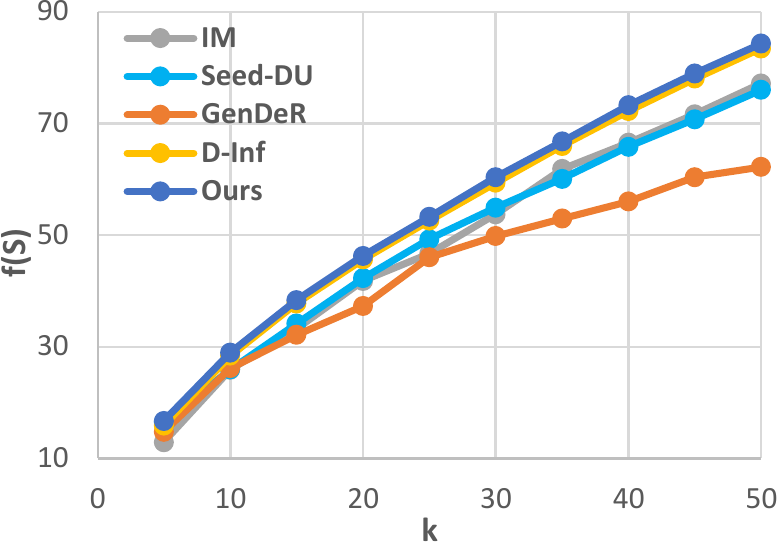}}
  \hspace{2ex}
  \subfigure[\textsc{Epinions}, CD]{
    \includegraphics[scale=0.43]{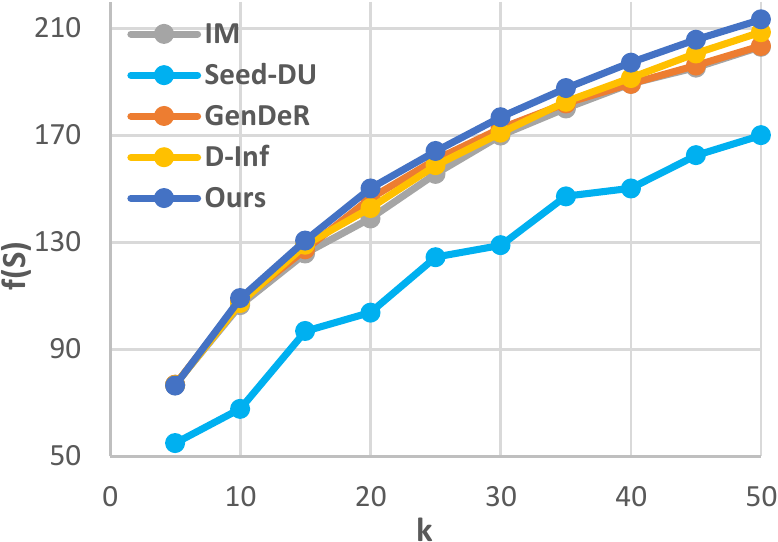}}
    \vspace{1mm}
  \caption{ADIM utility values on \textsc{FourArea} and \textsc{Epinions} with different utility functions (CES: CES ($\rho=1/2$), PC: Perfect Complements, CD: Cobb-Douglas).}
  \vspace{-1.2em}
  \label{fig:adim}
\end{figure}

\noindent \textbf{Algorithms.} The following algorithms are involved in our comparison: 

(1) \textbf{IM} \cite{kempe2003maximizing} is the traditional IM algorithm maximizing the spread over the whole network. 

(2) \textbf{GenDeR} \cite{he2012gender} is a generic diversified ranking algorithm. Here we use it for \textit{seed diversification}, where $\sigma(\{v\})$ is the ranking function and $\simm_C(u, v)$ is the similarity function. 

(3) \textbf{Seed-DU} \cite{tang2014diversified} is a \textit{seed-diversified} IM algorithm. Following \cite{tang2014diversified}, we set the diversity function $f(x)$ to be $\frac{x}{1+x}$. 

(4) \textbf{D-Inf} \cite{tang2014diversified} is an \textit{audience-diversified} IM algorithm. Still following \cite{tang2014diversified}, we set the diversity function to be $\frac{x}{1+x}$ and the balancing parameter $\gamma$ to be $0$.

(5) \textbf{Ours}, the framework proposed in this paper, uses \textsc{Greedy}/\textsc{Upper-Greedy} in ADIM and \textsc{Random-Greedy} in SDIM.

\vspace{1mm}

\noindent \textbf{Models and Parameters.} In SDIM, we set $a=b=1/2$ for Cobb-Douglas and $\beta = 0.05|V|$ for Perfect Substitutes and Perfect Complements. In ADIM, since Perfect Substitutes is too similar to traditional IM, we study CES ($\rho=1/2$) instead. For Perfect Complements and Cobb-Douglas, we set $\alpha_c = 1$ $(c=1,2,...,C)$. For CES ($\rho=1/2$), we set $\alpha_c = 1/C$ for normalization. We choose IC as our spreading model, where the activate probability $p_{uv}$ is $1/\text{deg}_\text{in}(v)$. 

\subsection{Utility Maximization Results}
\noindent \textbf{ADIM.} Figure \ref{fig:adim} shows the utility values of selected nodes in ADIM. We can observe that: (1) Ours consistently performs the best on both datasets with different utility functions. (2) In most cases, D-Inf performs the second best, whereas Seed-DU and GenDeR do not achieve satisfying utility values. This observation is aligned with their objectives. As we mentioned, D-Inf focuses on \textit{audience diversification} while Seed-DU and GenDeR consider to diversify \textit{seed nodes}. Although Tang et al. \cite{tang2014diversified} use \textit{homophily} to illustrate that diversified seeds may indicate diversified spreading results, both their experiments and ours show a gap between these two problem settings. (3) IM performs well with the CES utility ($\rho=1/2$). This finding indicates the similarity between CES ($\rho=1/2$) and traditional IM. In fact, traditional IM can be regarded as CES ($\rho=1$) when the whole network is a disjoint union of the $C$ communities. However, IM essentially ignores diversity. When it comes to Perfect Complements and Cobb-Douglas, IM performs significantly worse than Ours and D-Inf.

\vspace{1mm}

\begin{figure}[t]
\centering
  \subfigure[\textsc{FourArea}, PS]{
    \includegraphics[scale=0.42]{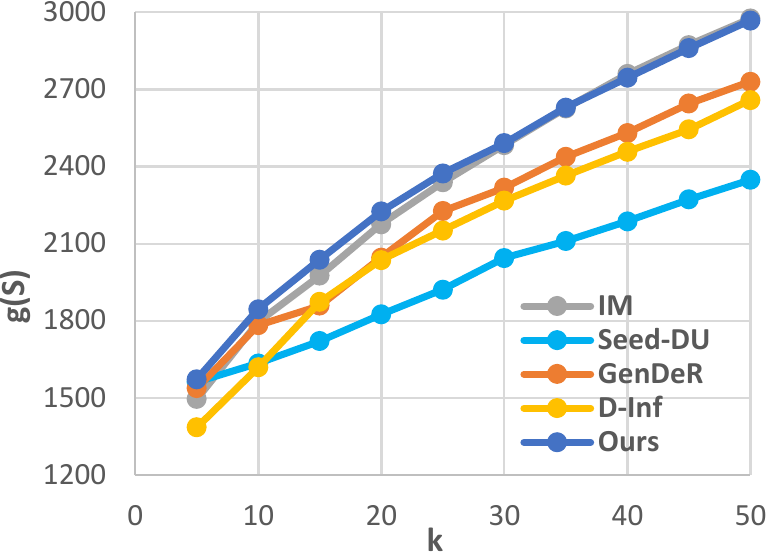}}
  \hspace{2ex}
  \subfigure[\textsc{Epinions}, PS]{
    \includegraphics[scale=0.42]{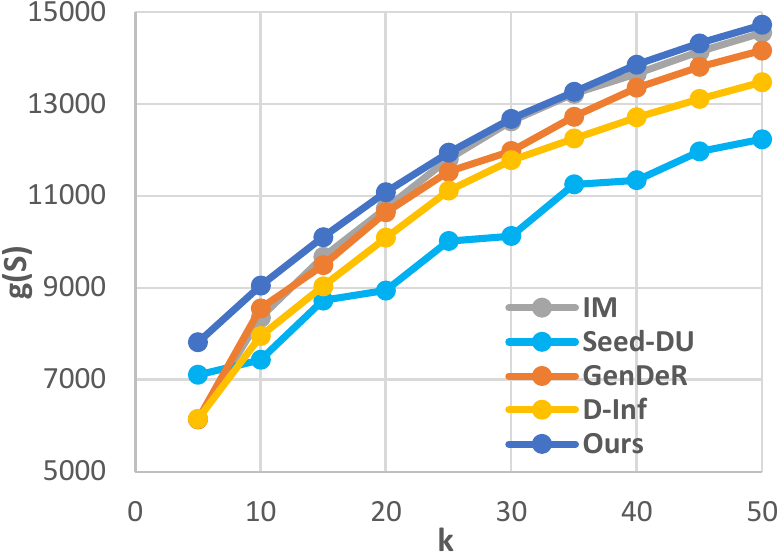}}
  \vspace{-2mm}
  \subfigure[\textsc{FourArea}, PC]{
    \includegraphics[scale=0.42]{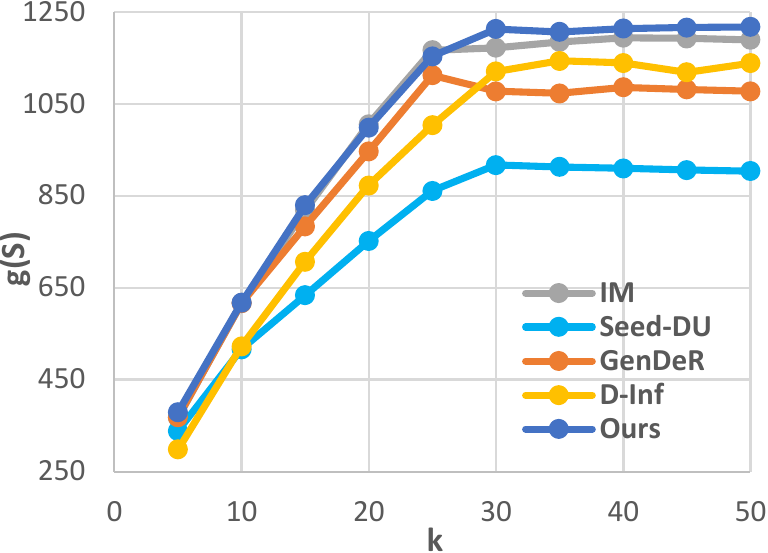}}
  \hspace{2ex}
  \subfigure[\textsc{Epinions}, PC]{
    \includegraphics[scale=0.42]{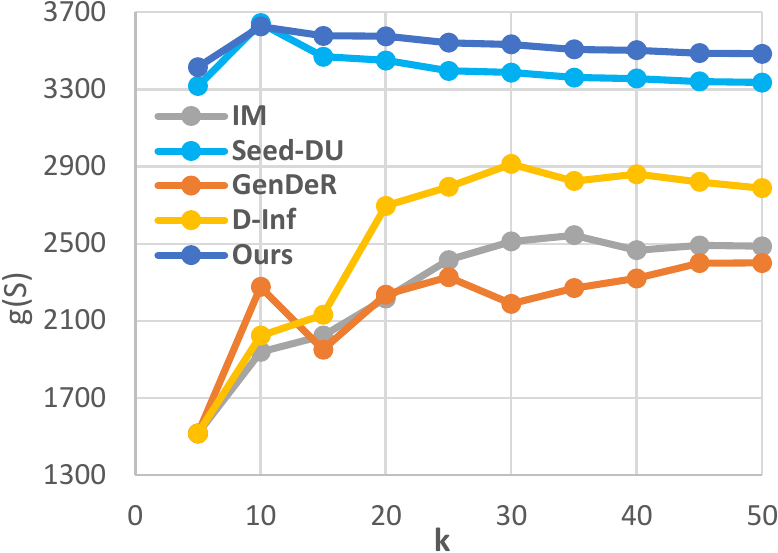}}
  \vspace{-2mm}
  \subfigure[\textsc{FourArea}, CD]{
    \includegraphics[scale=0.43]{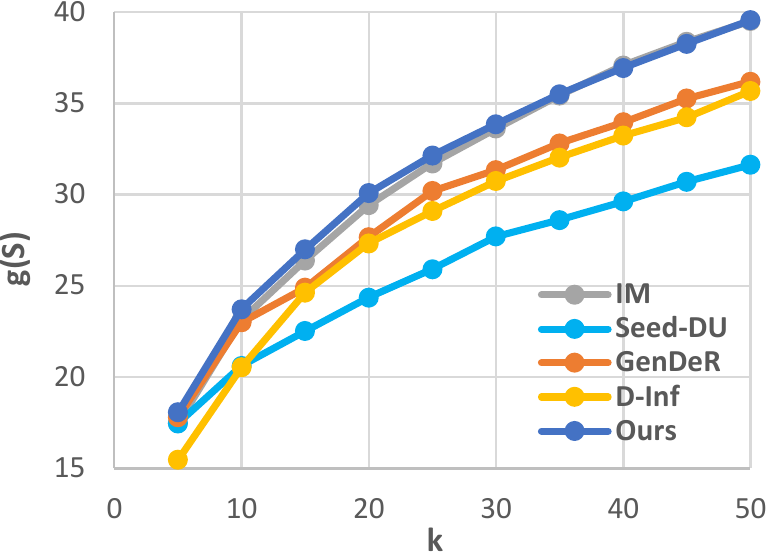}}
  \hspace{2ex}
  \subfigure[\textsc{Epinions}, CD]{
    \includegraphics[scale=0.43]{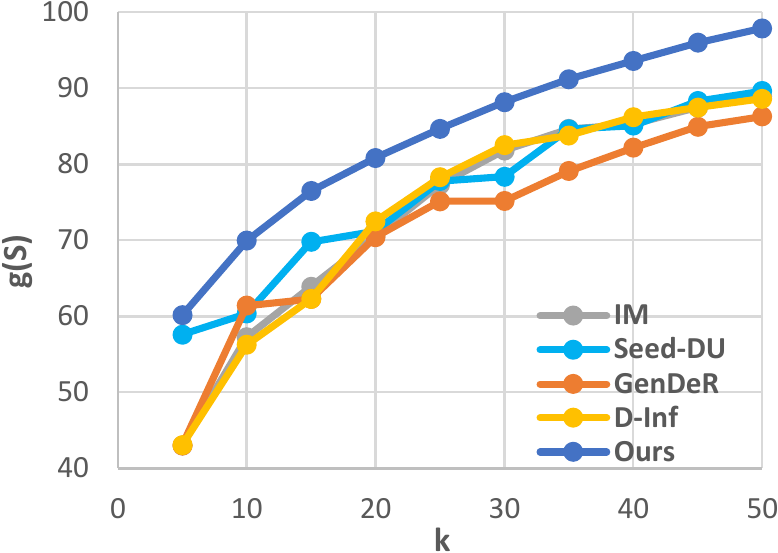}}
    \vspace{1mm}
  \caption{SDIM utility values on \textsc{FourArea} and \textsc{Epinions} with different utility functions (PS: Perfect Substitutes, PC: Perfect Complements, CD: Cobb-Douglas).}
   \vspace{-1.2em}
  \label{fig:sdimc}
\end{figure}

\noindent \textbf{SDIM.} We use $\simm_C(\cdot,\cdot)$ defined in Section \ref{sec:sdim} as our similarity function. Figures \ref{fig:sdimc} shows the utilities of selected nodes in SDIM. 
We have the following findings: (1) Again, Ours consistently performs the best. When using Perfect Complements and Cobb-Douglas utilities, we can outperform the baselines by a large margin; when using Perfect Substitutes, we are still the best, but the advantage against IM is slight. We will explain the reason Section \ref{sec:explanation}.
(2) In SDIM, GenDeR and Seed-DU have competitive performances with D-Inf. For example, GenDeR achieves higher utility values in most cases when using Perfect Substitutes and Cobb-Douglas utilities, and Seed-DU can outperform D-Inf significantly with the Perfect Complements utility on \textsc{Epinions}.

\subsection{Audience Diversification Results} 
Higher utilities are good news, but do not necessarily indicate satisfying results. Putting the utilities aside, we would like to prove that our algorithms can really diversify the activated nodes without hurting the spread in ADIM.

\vspace{1mm}

\noindent \textbf{Evaluation Metrics.} Following \cite{tang2014diversified}, we define the following two metrics.
\begin{equation}
\begin{split}
&\text{Entropy}(S) = \sum_{i=1}^C -p_i\log p_i, \text{\ \ where\ } p_i = \frac{\sigma_i(S)}{\sum_{i=1}^C\sigma_i(S)}.\\
&\text{Spread}(S) = \mathbb{E}[|S_{\tt act} \cap (V_1\cup...\cup V_C)|]. \notag
\end{split}
\end{equation}
Intuitively, $p_i$ can be interpreted as the proportion of influence distributed to community $V_i$, and Entropy$(S)$ reflects the degree of balance with respect to the influence spread. Spread$(S)$, from an orthogonal perspective, measures how many users \textit{in target communities} are affected. (There are nodes not belonging to any target community $V_i$.)

\vspace{1mm}

\noindent \textbf{Results.}
Tables \ref{tab:efour} and \ref{tab:eepin} show the Entropy and Spread of cascading results on \textsc{FourArea} and \textsc{Epinions} when $k = 50$. Here ``Ours-CES'' means we select top-$k$ nodes using our approach (i.e., \textsc{Greedy}/\textsc{Upper-Greedy}) with CES ($\rho=1/2$). Similar meanings can be inferred for ``Ours-PC'' and ``Ours-CD''. For each algorithm, we calculate its percentage increase/decrease in comparison with IM. On the one hand, when we only focus on Entropy, it can be observed that Ours-PC performs the best on \textsc{FourArea} and the second best (and on par with the best) on \textsc{Epinions}. This indicates PC is the most applicable utility when users intend to put more emphasis on diversity. On the other hand, in many practical scenarios of IM, our goal is to increase the diversity of audience without hurting influence spread. Among all compared methods, Seed-DU and D-Inf sacrifice Spread during diversification; GenDeR does not diversify the results; only Ours-CES and Ours-CD increase Entropy and Spread simultaneously.

\subsection{Seed Diversification Results}
Following the same way, in SDIM, we would like to prove our success in seed diversification from the perspectives other than utility values. Following the evaluation metrics in \cite{zhu2007improving,mei2010divrank,he2012gender}, we conduct experiments on an actor professional network.

\vspace{1mm}

\noindent \textbf{Dataset.} The \textsc{Imdb} network is constructed from the Internet Movie Database. Each actor/actress is represented by a node, and the edges between two nodes denote their co-starred movies. Unseen by the algorithms, each actor/actress is associated with a country. The dataset we use \footnote{www.kaggle.com/carolzhangdc/imdb-5000-movie-dataset} involves 5,044 movies and 6,271 actors/actresses, generating an undirected network with 15,060 edges.

\vspace{1mm}

\noindent \textbf{Evaluation Metrics.} Zhu et al. \cite{zhu2007improving} propose two diversity measures in a particular context of ranking movie stars, i.e., \textit{Country Coverage} and \textit{Movie Coverage}, which are the number of distinct countries and movies associated with the selected actors/actresses. Previous studies \cite{zhu2007improving,mei2010divrank,he2012gender} expect that higher coverages indicate more influential and more diverse results.

\begin{table}[t]
\footnotesize
\centering
\caption{Entropy and Spread of cascading results ($k=50$) on \textsc{FourArea}. For each algorithm, we calculate its percentage increase/decrease in comparison with \textbf{IM}. (CES: CES ($\rho=1/2$), PC: Perfect Complements, CD: Cobb-Douglas.)}
\vspace{-0.5em}
\scalebox{0.77}{
	\begin{tabular}{c|ccccccc}
		\hline
		Method                   & \textbf{IM}      & \textbf{GenDeR}  & \textbf{Seed-DU} & \textbf{D-Inf}   & \textbf{Ours-CES}     & \textbf{Ours-PC}      & \textbf{Ours-CD}      \\ \hline
		\multirow{2}{*}{Entropy} & 1.883   & 1.628   & 1.947   & 1.977   & 1.923   & 1.994   & 1.943   \\
		& -       & \color{red}{\bf-13.5\%} & \color{blue}{\bf+3.4\%}  & \color{blue}{\bf+5.0\%}  & \color{blue}{\bf+2.1\%}  & \color{blue}{\bf+5.8\%}  & \color{blue}{\bf+3.2\%}  \\ \hline
		\multirow{2}{*}{Spread}  & 344.20 & 338.54 & 315.97 & 339.24 & 354.31 & 256.14 & 350.90 \\
		& -       & \color{red}{\bf -1.6\%}  & \color{red}{\bf -8.2\%}  & \color{red}{\bf -1.4\%}  & \color{blue}{\bf+2.9\%}  & \color{red}{\bf -25.6\%} & \color{blue}{\bf+1.9\%}  \\ \hline
	\end{tabular}
	}
	\vspace{-1em}
	\label{tab:efour}
\end{table}

\begin{table}[t]
	\footnotesize
	\caption{Entropy and Spread of cascading results ($k=50$) on \textsc{Epinions}. For each algorithm, we calculate its percentage increase/decrease in comparison with \textbf{IM}. (CES: CES ($\rho=1/2$), PC: Perfect Complements, CD: Cobb-Douglas.)}
	\vspace{-0.5em}
\scalebox{0.77}{
	\begin{tabular}{c|ccccccc}
		\hline
		Method                   & \textbf{IM}      & \textbf{GenDeR}  & \textbf{Seed-DU} & \textbf{D-Inf}   & \textbf{Ours-CES}     & \textbf{Ours-PC}      & \textbf{Ours-CD}      \\ \hline
		\multirow{2}{*}{Entropy} & 2.774    & 2.769    & 2.833    & 2.820    & 2.793    & 2.830    & 2.806    \\
		& -        & \color{red}{\bf-0.2\%}   & \color{blue}{\bf+2.1\%}   & \color{blue}{\bf+1.7\%}   & \color{blue}{\bf+0.7\%}   & \color{blue}{\bf+2.0\%}   & \color{blue}{\bf+1.2\%}   \\ \hline
		\multirow{2}{*}{Spread}  & 3486.2 & 3503.9 & 2701.8 & 3305.8 & 3549.3 & 3017.0 & 3487.0 \\
		& -        & \color{blue}{\bf+0.5\%}   & \color{red}{\bf-22.5\%}  & \color{red}{\bf-5.2\%}   & \color{blue}{\bf+1.8\%}   & \color{red}{\bf-13.5\%}  & \color{blue}{\bf+0.02\%}  \\ \hline
	\end{tabular}
	\label{tab:eepin}
}
\end{table}

\begin{figure}[!t]
\centering
  \subfigure[Country Coverage]{
    \includegraphics[scale=0.42]{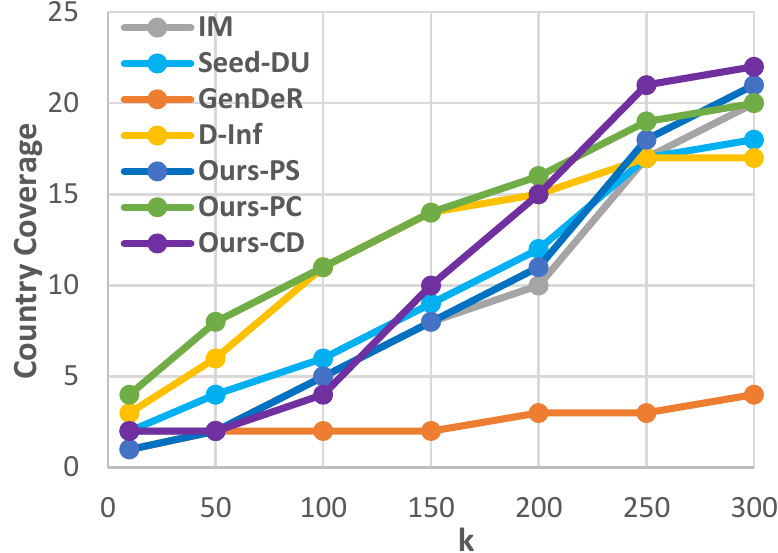}}
  \hspace{2ex}
  \subfigure[Movie Coverage]{
    \includegraphics[scale=0.42]{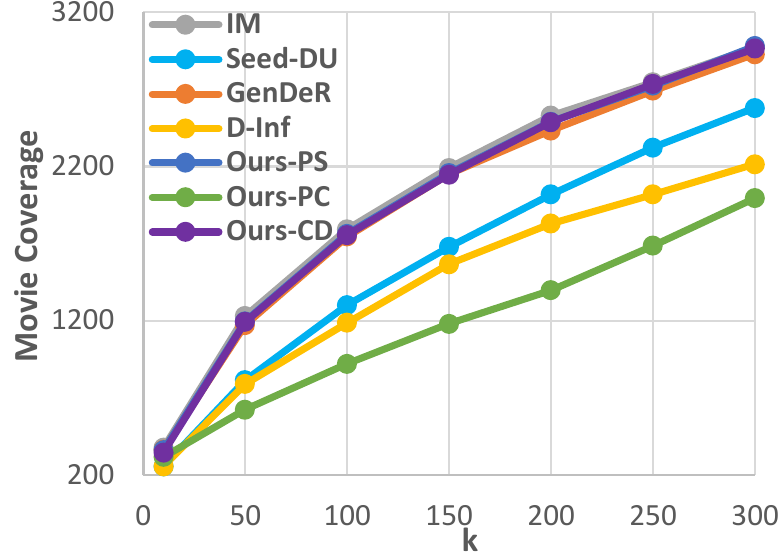}}
  \caption{Country Coverage and Movie Coverage on the \textsc{Imdb} network (PS: Perfect Substitutes, PC: Perfect Complements, CD: Cobb-Douglas).}
  \vspace{-1.2em}
  \label{fig:imdb}
\end{figure}

\vspace{1mm}

\noindent \textbf{Results.} The results are shown in Figure \ref{fig:imdb}. Intuitively, Country Coverage mainly evaluates seed diversity (if we treat each country as a community, covering more countries essentially means the seeds are extracted from more communities) while Movie Coverage cares more about influence power (if an actor/actress appears in more movies, he/she will co-star with more performers and should be more familiar to the audience as well). In Figure \ref{fig:imdb}(b), Seed-DU, D-Inf and Ours-PC perform evidently worse, and all the other methods are on par with each other. Meanwhile, in Figure \ref{fig:imdb}(a), Ours-PC and D-Inf are the best two when $k$ is small. Without them, Ours-CD gives the most diversified results. Similar to audience diversification, we explain these observations from two perspectives. On the one hand, when we are more willing to substitute spread for diversity, Ours-PC can give us the most diversified results. On the other hand, if we would like to diversify the results with little reduce in influence power, Ours-CD is the best choice. 

\subsection{Explanation of the Results}
\label{sec:explanation}
\begin{figure}[t]
\centering
    \includegraphics[scale=0.35]{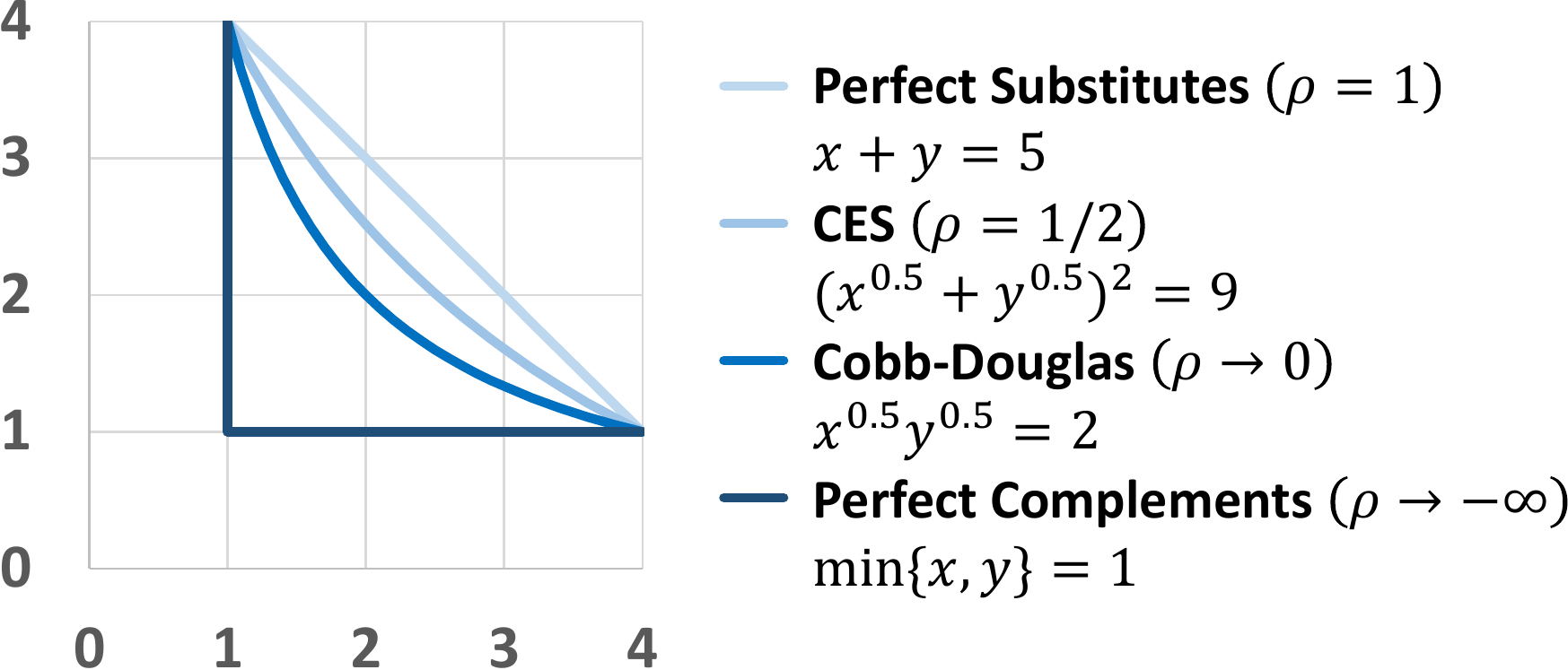}
    \vspace{-0.5em}
  \caption{Indifference curves of Perfect Substitutes, CES ($\rho=1/2$), Cobb-Douglas and Perfect Complements.}
  \vspace{-1.2em}
  \label{fig:util}
\end{figure}
We have many observations in the experiments: In SDIM, Ours can only outperform IM slightly with the PS utility. In audience diversification, Ours-PC has the highest Entropy among the three utilities, while Ours-CES has the highest Spread and Ours-CD ranks the second in both metrics. In seed diversification, Ours-PC has the highest Country Coverage among the three utilities, followed by Ours-CD and then Ours-PS.
These findings can be explained by the properties of the utilities. In fact, CES has three popular special cases, which are exactly PS ($\rho= 1$), CD ($\rho\rightarrow 0$) and PC ($\rho\rightarrow -\infty$). We plot the indifference curves of different utility functions in Figure \ref{fig:util}. We can observe that the smaller the $\rho$ is, the more convex the curve is to the origin, in which case we are more willing to substitute spread for diversity. From this perspective, PC emphasizes the most on diversity, followed by CD, CES ($0<\rho<1$), and then PS.

\section{Conclusion}
We have presented an IM framework that works for both audience diversification and seed diversification. We formulate the ADIM and SDIM tasks by carefully designing the objective to jointly describe influence spread and diversity. Three economic utilities with nice properties are adopted. Theoretically, we present various approximation algorithms (\textsc{Greedy}, \textsc{Upper-Greedy} and \textsc{Random-Greedy}) to maximize the utilities. Practically, we validate the effectiveness of our solutions in both utility maximization and audience/seed diversification. There are still open issues in light of these results. First, it would be interesting to devise more efficient versions of the proposed algorithms. Second, more effort is needed to explore a comprehensive metric jointly evaluating spread and diversity.

\bibliographystyle{abbrv}
\bibliography{sigir17}

\appendix

\subsection{Proof of Theorem 1}
\begin{proof}
(P1) is trivial. We only prove (P2).

\vspace{1mm}

\noindent \textbf{CES ($0<\rho\leq 1$).} We know $x^\rho$ is concave. Using Jensen's inequality, we have
$$
\sum_{i=1}^k x_i^\rho \leq k\Big(\frac{x_1+...+x_k}{k}\Big)^\rho.
$$
Therefore,
$$
\sum_{i=1}^C x_i^\rho \leq k\Big(\frac{x_1+...+x_k}{k}\Big)^\rho + \sum_{i=k+1}^Cx_i^\rho.
$$

\vspace{1mm}

\noindent \textbf{Perfect Complements.} We have
$$
\min_{1\leq i \leq k}x_i \leq \frac{x_1+...+x_k}{k}.
$$
Therefore,
\begin{equation}
\begin{split}
\min_{1\leq i \leq C}x_i &= \min\{\min_{1\leq i \leq k}x_i, x_{k+1}, ..., x_C\} \\
& \leq \min\{\frac{x_1+...+x_k}{k}, x_{k+1}, ..., x_C\}. \notag
\end{split}
\end{equation}

\vspace{1mm}

\noindent \textbf{Cobb-Douglas.} Since the arithmetic mean is always greater than or equal to the geometric mean, we have
$$
\Big(\frac{x_1+...+x_k}{k}\Big)^k \geq \prod_{i=1}^k x_i.
$$
Therefore,
\begin{equation}
\begin{split}
\Big(\prod_{i=1}^C x_i\Big)^{1/C} &= \Big(\prod_{i=1}^k x_i \cdot \prod_{i=k+1}^C x_i\Big)^{1/C} \\ 
&\leq \Big(\Big(\frac{x_1+...+x_k}{k}\Big)^{k}\cdot \prod_{i=k+1}^C x_i\Big)^{1/C}. \notag
\end{split}
\end{equation}
\end{proof}
\subsection{Proof of Theorem 2}
\begin{proof}
(P1) is trivial. We only prove (P3).

\vspace{1mm}

\noindent \textbf{Perfect Substitutes.} We have
$$
x_1+x_2 = G_S(x_1, x_2) = G_S(x_1-\epsilon, x_2+\delta) = x_1+x_2+\delta-\epsilon.
$$
Therefore, $\delta = \epsilon$. Then
$$
G_S(x_1-2\epsilon, x_2+2\delta) = x_1+x_2 = G_S(x_1-\epsilon, x_2+\delta).
$$

\vspace{1mm}

\noindent \textbf{Perfect Complements.}

\underline{\textit{Case 1.}} If $x_1 \leq x_2$, then $x_1-\epsilon \leq x_2 + \delta$. Therefore,
$$
x_1 = G_C(x_1, x_2) = G_C(x_1-\epsilon, x_2+\delta) = x_1-\epsilon.
$$
We have $\epsilon = 0$, and then 
$$
G_C(x_1-2\epsilon, x_2+2\delta) = x_1-2\epsilon = x_1-\epsilon = G_C(x_1-\epsilon, x_2+\delta).
$$

\underline{\textit{Case 2.}} If $x_1 > x_2$ and $x_1-\epsilon \geq x_2+\delta$, then
$$
x_2 = G_C(x_1, x_2) = G_C(x_1-\epsilon, x_2+\delta) = x_2+\delta.
$$
We have $\delta = 0$, and then 
$$
G_C(x_1-2\epsilon, x_2+2\delta) \leq x_2+2\delta = x_2+\delta = G_C(x_1-\epsilon, x_2+\delta).
$$

\underline{\textit{Case 3.}} If $x_1 > x_2$ and $x_1-\epsilon < x_2+\delta$, then
$$
x_2 = G_C(x_1, x_2) = G_C(x_1-\epsilon, x_2+\delta) = x_1-\epsilon.
$$
Therefore, $x_1-2\epsilon \leq x_2$, and we have
\begin{equation}
\begin{split}
G_C(x_1-2\epsilon, x_2+2\delta) & \leq x_1-2\epsilon \leq x_2 \\
& = G_C(x_1, x_2)=G_C(x_1-\epsilon, x_2+\delta). \notag
\end{split}
\end{equation}

\vspace{1mm}

\noindent \textbf{Cobb-Douglas.} We have
$$
1 = \frac{G_D(x_1-\epsilon, x_2+\delta)}{G_D(x_1, x_2)} = \frac{(x_1-\epsilon)^\alpha(x_2+\delta)^\beta}{x_1^\alpha x_2^\beta},
$$
and
$$
\frac{G_D(x_1-2\epsilon, x_2+2\delta)}{G_D(x_1-\epsilon, x_2+\delta)} = \frac{(x_1-2\epsilon)^\alpha(x_2+2\delta)^\beta}{(x_1-\epsilon)^\alpha(x_2+\delta)^\beta}.
$$

Note that
$$
\frac{x_1-2\epsilon}{x_1-\epsilon} = 1 - \frac{\epsilon}{x_1-\epsilon} \leq 1 - \frac{\epsilon}{x_1} = \frac{x_1-\epsilon}{x_1},
$$
and
$$
\frac{x_2+2\delta}{x_2+\delta} = 1 + \frac{\delta}{x_2+\delta} \leq 1 + \frac{\delta}{x_2} = \frac{x_2+\delta}{x_2}.
$$
Therefore,
\begin{equation}
\begin{split}
\frac{G_D(x_1-2\epsilon, x_2+2\delta)}{G_D(x_1-\epsilon, x_2+\delta)} &= \Big(\frac{x_1-2\epsilon}{x_1-\epsilon}\Big)^\alpha \Big(\frac{x_2+2\delta}{x_2+\delta}\Big)^\beta \\ 
&\leq \Big(\frac{x_1-\epsilon}{x_1}\Big)^\alpha \Big(\frac{x_2+\delta}{x_2}\Big)^\beta \\
&=\frac{G_D(x_1-\epsilon, x_2+\delta)}{G_D(x_1, x_2)} = 1. \notag
\end{split}
\end{equation}
\end{proof}
\subsection{Proof of Theorem 3}
\begin{proof}
We have $f_{CES}(S)^\rho = \sum_{c=1}^C (\alpha_c \sigma_c(S))^\rho$. Note that $\alpha_c \sigma_c(S)$ is monotonic and submodular and $x^\rho$ is non-decreasing and concave. Using Theorem 1 in \cite{lin2011class}, their composition $(\alpha_c \sigma_c(S))^\rho$ is also monotonic and submodular. Therefore, $f_{CES}(S)^\rho$ is monotonic and submodular.

Suppose the optimal solution of $g_{CES}$ is $S_{CES}^*$. According to \cite{nemhauser1978analysis}, {\sc Greedy}($f_{CES}^\rho,k$) guarantees that
$$
f_{CES}(S)^\rho \geq (1-\frac{1}{e}-\epsilon)f_{CES}(S_{CES}^*)^\rho.
$$
Thus,
$$
f_{CES}(S) \geq (1-\frac{1}{e}-\epsilon)^{1/\rho} f_{CES}(S_{CES}^*).
$$

When $\rho = 1$, we have the common $(1-1/e-\epsilon)$ approximation ratio for Perfect Substitutes.
\end{proof}

\subsection{Proof of Theorem 4}
\begin{proof}
Our goal is to prove that $\forall \epsilon >0$, the following two problems are \texttt{NP}-hard. 

(1) Find a set $S$ $(|S|\leq k)$ such that $f_C(S) \geq \epsilon \cdot f_C(S_C^*)$, where $S_C^*$ is the optimal solution for Perfect Complements. 

(2) Find a set $S$ $(|S|\leq k)$ such that $f_D(S) \geq \epsilon\cdot f_D(S_D^*)$, where $S_D^*$ is the optimal solution for Cobb-Douglas.

Consider the Vertex-Cover problem: Given a graph $G=(V,E)$, we need to determine whether there is a node set $|S_V|\leq k$ such that for any node in $V$, at least one of its neighbor (or itself) is in $S_V$. This is a famous NP-hard problem.

Given a Vertex-Cover instance $G=(V,E)$ (where $V=\{v_1,...,v_n\}$), we construct a bipartite graph $G' = (V', E')$, where $V' = \{u_{11},...,u_{1n},$ $u_{21},...,u_{2n}\}$. There is a directed edge from $u_{1i}$ to $u_{2j}$ if and only if $(v_i,v_j)\in E$ or $i=j$. The activating probability is 1 for all the edges. Consider the ADIM problem on $G'$. We have $n$ communities in total, where $V_c = \{u_{2c}\}$ $(c=1,2,...,n)$.

Assume there is a vertex cover $|S_V| = \{v_{a}|a\in A\}$ $(|A|\leq k)$. Let $S = \{u_{1a}|a\in A\}$. Since $|S_V|$ is a vertex cover, all of the $u_{2c}$ $(c=1,2,...,n)$ will be activated. Therefore, $f_C(S) = \min_c \alpha_c > 0$, and $f_D(S) = 1 > 0$.

On the other side, assume there is no vertex cover of size $k$. Consider a seed set $|S| = \{u_{1a},u_{2b}|a\in A, b\in B\}$. First, let $S' = \{u_{1a}|a\in A \cup B\}$. It is obvious that $S'$ can activate all the nodes activated by $S$. However, if $|S'|\leq k$, at least one $u_{2c}$ cannot be activated by $S'$ (and $S$). Therefore, $f_C(S) = f_D(S) = 0$.

According to the analysis above, we know that there is a vertex cover of size $k$ if and only if there is a seed set $S$ with $f_C(S) >0$. If there is an algorithm which guarantees $f_C(S) \geq \epsilon\cdot f_C(S_C^*)$, consider the output $\Tilde{S}$ of this algorithm.

If $f_C(\Tilde{S}) > 0$, then there is a vertex cover of size $k$.

If $f_C(\Tilde{S}) = 0$, then $\forall S$ with $|S|\leq k$, $f_C(S) \leq f_C(S_C^*) \leq \frac{1}{\epsilon}f_C(\Tilde{S}) = 0$. There is no vertex cover of size $k$.

In summary, the algorithm can also judge the existence of a vertex cover, which finishes our reduction. For $f_D(\cdot)$, we can follow the same way.
\end{proof}

\subsection{Proof of Theorem 5}
\begin{proof}
For any community $V_i$, we know that $\alpha_i\sigma_i(\cdot)$ is monotonic and submodular. Therefore, when we use \textsc{Greedy} to maximize $\alpha_i\sigma_i(\cdot)$, we have $\alpha_i\sigma_i(S_i) \geq (1-1/e-\epsilon)\alpha_i\sigma_i(S_{i}^*)$, where $S_{i}^*$ is the optimal solution for $\sigma_i(\cdot)$. Since $\alpha_i\sigma_i(S)\geq f_C(S)$ for any $S$, $\alpha_i\sigma_i(S_{i}^*) \geq \alpha_i\sigma_i(S_{C}^*) \geq f_{C}(S_{C}^*)$. Therefore,
\begin{equation}
\label{eqn:upper}
\begin{split}
 f_C(S) &\geq \frac{f_C(S_i)}{\alpha_i\sigma_i(S_i)} \alpha_i\sigma_i(S_i) \\
        &\geq \frac{f_C(S_i)}{\alpha_i\sigma_i(S_i)} (1-\frac{1}{e}-\epsilon)\alpha_i\sigma_i(S_{i}^*) \\
        &\geq \frac{f_C(S_i)}{\alpha_i\sigma_i(S_i)} (1-\frac{1}{e}-\epsilon)f_{C}(S_{C}^*).
\end{split}
\end{equation}

Inequality (\ref{eqn:upper}) holds for any $i$. Therefore, 
\begin{equation}
f_C(S) \geq \max_{1\leq i \leq C}\frac{f_C(S_i)}{\alpha_i\sigma_i(S_i)}(1-\frac{1}{e}-\epsilon)f_C(S_C^*). \notag
\end{equation}
\end{proof}

\subsection{Proof of Theorem 6}
\begin{proof}
Using Lemma 1, we know that $f_{D+}(S)$ is monotonic and submodular. Similar to Theorem 5, we have $f_{D+}(S_+)\geq (1-1/e-\epsilon)f_{D+}(S_{D+}^*)$, where $S_{D+}^*$ is the optimal solution for $f_{D+}(\cdot)$. Therefore,
\begin{equation}
\begin{split}
 f_D(S) &\geq \frac{f_D(S_+)}{f_{D+}(S_+)} f_{D+}(S_+) \\
        &\geq \frac{f_D(S_+)}{f_{D+}(S_+)} (1-\frac{1}{e}-\epsilon) f_{D+}(S_{D+}^*) \\
        &\geq \frac{f_D(S_+)}{f_{D+}(S_+)} (1-\frac{1}{e}-\epsilon) f_{D}(S_{D}^*). \notag
\end{split}
\end{equation}
\end{proof}

\subsection{Proof of Lemma 2}
\begin{proof}
\textbf{Non-negativity.} We have 
$$
\sum_{u,v \in S, u\neq v}\simm(u,v) \leq \sum_{u,v \in S, u\neq v} 1 \ =\ k(k-1).
$$
Therefore, $\tilde{d}(S) = 1 - \frac{1}{k(k-1)}\sum_{u,v \in S, u\neq v}\simm(u,v) \geq 0$.

\vspace{1mm}

\noindent \textbf{(Decreasing) Monotonicity.} Since $\simm(u, v) \geq 0$, $\forall S \subseteq T$,
$$
\sum_{u,v \in S, u\neq v}\simm(u,v) \leq \sum_{u,v \in T, u\neq v}\simm(u,v)
$$
Therefore, $\tilde{d}(S) \geq \tilde{d}(T)$

\vspace{1mm}

\noindent \textbf{Submodularity.} $\forall S \subseteq T$ and $x \notin T$,
\begin{equation}
\begin{split}
    & \sum_{u,v \in S\cup\{x\}, u\neq v}\simm(u,v) - \sum_{u,v \in S, u\neq v}\simm(u,v) \\
    =&\ \ \ \ \ \ \ \ \ \sum_{u\in S}\simm(u,x) + \sum_{u\in S}\simm(x,u)\\
\leq &\ \ \ \ \ \ \ \ \ \sum_{u\in T}\simm(u,x) + \sum_{u\in T}\simm(x,u)\\
    =&\sum_{u,v \in T\cup\{x\}, u\neq v}\simm(u,v) - \sum_{u,v \in T, u\neq v}\simm(u,v). \notag
\end{split}
\end{equation}
Therefore, $\sum_{u,v \in S, u\neq v}\simm(u,v)$ is supermodular, indicating $\tilde{d}(S) = 1 - \frac{1}{k(k-1)}\sum_{u,v \in S, u\neq v}\simm(u,v)$ is submodular.
\end{proof}

\subsection{Proof of Theorem 7}
\begin{proof}
\noindent \textbf{Perfect Substitutes.} $\tilde{g}_S(S)$ is the sum of two submodular functions, which is also submodular.

\vspace{1mm}

\noindent \textbf{Perfect Complements.} $\forall S, T$, we will prove that $\tilde{g}_C(S) + \tilde{g}_C(T)\geq \tilde{g}_C(S\cup T) + \tilde{g}_C(S\cap T)$, which is an equivalent definition for submodular functions \cite{nemhauser1978analysis}.

\underline{\textit{Case 1.}} If $\tilde{g}_C(S) = \sigma(S)$ and $\tilde{g}_C(T) = \sigma(T)$ (which means $\sigma(\cdot)$ is smaller than $\beta \tilde{d}(\cdot)$ for both $S$ and $T$), then 
\begin{equation}
\begin{split}
    &\ \tilde{g}_C(S) + \tilde{g}_C(T) \\
    = &\ \sigma(S) + \sigma(T) \\
    \geq& \ \sigma(S\cup T) + \sigma(S \cap T) \\
    \geq& \ \min\{\sigma(S\cup T), \beta \tilde{d}(S\cup T)\} + \min\{\sigma(S\cap T), \beta \tilde{d}(S\cap T)\} \\
    =&\  \tilde{g}_C(S\cup T) + \tilde{g}_C(S\cap T). \notag
\end{split}
\end{equation}

\underline{\textit{Case 2.}} If $\tilde{g}_C(S) = \beta \tilde{d}(S)$ and $\tilde{g}_C(T) = \beta \tilde{d}(T)$, we have the proof similar to that of Case 1.

\underline{\textit{Case 3.}} If $\tilde{g}_C(S) = \sigma(S)$ and $\tilde{g}_C(T) = \beta \tilde{d}(T)$, then
\begin{equation}
\begin{split}
    &\ \tilde{g}_C(S) + \tilde{g}_C(T) \\
    = &\ \sigma(S) + \beta \tilde{d}(T) \\
    \geq& \ \sigma(S\cup T) + \sigma(S \cap T) - \sigma(T) + \beta \tilde{d}(T) \\
    \geq& \ \sigma(S\cup T) + \sigma(S \cap T) - \sigma(S\cap T) + \beta \tilde{d}(S\cap T) \\
    & \textrm{(note that $\sigma(\cdot)$ is increasing and $\beta \tilde{d}(\cdot)$ is decreasing.)} \\
    =& \ \sigma(S\cup T) + \beta \tilde{d}(S\cap T) \\
    \geq&\  \tilde{g}_C(S\cup T) + \tilde{g}_C(S\cap T). \notag
\end{split}
\end{equation}

\underline{\textit{Case 4.}} If $\tilde{g}_C(S) = \beta \tilde{d}(S)$ and $\tilde{g}_C(T) = \sigma(T)$, we have the proof similar to that of Case 3.

\vspace{1mm}

\noindent \textbf{Cobb-Douglas.} We prove $\tilde{g}_D(S)$ is submodular for any $0 \leq a, b \leq 1$ (which means $a+b=1$ is not needed here). We first deal with the case where $a=b=1$.
According to the properties of $\sigma(\cdot)$ and $\tilde{d}(\cdot)$, $\forall S \subseteq T$ and $x \notin T$, we can have the following notations:
\begin{equation}
\begin{split}
    &\sigma(S\cup\{x\}) - \sigma(S) = \Delta\sigma,\  \sigma(T\cup\{x\}) - \sigma(T) = \Delta\sigma - \epsilon_1, \\
    & \tilde{d}(S\cup\{x\}) - \tilde{d}(S) = -\Delta d,\  \tilde{d}(T\cup\{x\}) - \tilde{d}(T) = -\Delta d - \epsilon_2, \notag
\end{split}
\end{equation}
where $\Delta d \geq 0$, $\epsilon_2 \geq 0$ and $\Delta\sigma \geq \epsilon_1 \geq 0$.

Now we have
\begin{equation}
\begin{split}
    & \tilde{g}_D(T\cup\{x\}) - \tilde{g}_D(T) \\
    =&\ \tilde{d}(T\cup\{x\})\cdot \sigma(T\cup\{x\}) - \tilde{d}(T)\cdot\sigma(T) \\
    =&\ (\tilde{d}(T)-\Delta d - \epsilon_2)(\sigma(T)+\Delta\sigma - \epsilon_1) - \tilde{d}(T)\cdot\sigma(T) \\
    =&\ (\Delta\sigma - \epsilon_1)\tilde{d}(T) - (\Delta d + \epsilon_2)\sigma(T) - (\Delta\sigma - \epsilon_1)(\Delta d + \epsilon_2) \\
    \leq & \ (\Delta\sigma - \epsilon_1)\tilde{d}(S) - (\Delta d + \epsilon_2)\sigma(S) - (\Delta\sigma - \epsilon_1)(\Delta d + \epsilon_2) \\
        & \textrm{(note that $\sigma(\cdot)$ is increasing and $\tilde{d}(\cdot)$ is decreasing.)} \\
    =&\ \Delta\sigma\cdot \tilde{d}(S) - \Delta d\cdot\sigma(S) - \Delta\sigma\Delta d \\
     &\ -\epsilon_1(\tilde{d}(S)-\Delta d) - \epsilon_2\sigma(S) - \epsilon_2(\Delta\sigma - \epsilon_1) \\
    \leq & \ \Delta\sigma\cdot \tilde{d}(S) - \Delta d\cdot\sigma(S) - \Delta\sigma\Delta d \\
    =&\ \tilde{g}_D(S\cup\{x\}) - \tilde{g}_D(S). \notag
\end{split}
\end{equation}

For other cases, we just need to prove that $\sigma(\cdot)^a$ is non-negative, increasing and submodular, and $\tilde{d}(\cdot)^b$ is non-negative, decreasing and submodular, then we can follow the same way as above to prove the submodularity of their product. 

For $\sigma(S)^a$, non-negativity and monotonicity are trivial. As for submodularity, we know $\sigma(S)$ is monotonic and submodular and $x^a\ (0\leq a \leq 1)$ is non-decreasing and concave. Using Theorem 1 in \cite{lin2011class}, their composition $\sigma(S)^a$ is also submodular.

For $\tilde{d}(S)^b$, the proof is similar.
$\forall S \subseteq T$ and $x \notin T$,

\underline{\textit{Case 1.}} If $\tilde{d}(T)\leq \tilde{d}(S\cup\{x\})$, we have 
\begin{equation}
\begin{split}
     \tilde{d}(T)^b - \tilde{d}(T\cup\{x\})^b
    =\ b(\xi_1)^{b-1} \cdot (\tilde{d}(T) - \tilde{d}(T\cup\{x\})) \notag
\end{split}
\end{equation}
according to Lagrange's mean value theorem. Similarly,
\begin{equation}
\begin{split}
     \tilde{d}(S)^b - \tilde{d}(S\cup\{x\})^b
    =\ b(\xi_2)^{b-1} \cdot (\tilde{d}(S) - \tilde{d}(S\cup\{x\})) \notag
\end{split}
\end{equation}

We know $\tilde{d}(\cdot)$ is decreasing and submodular. Therefore,
$$
\tilde{d}(T) - \tilde{d}(T\cup\{x\}) \geq \tilde{d}(S) - \tilde{d}(S\cup\{x\}) \geq 0.
$$
Besides, because $\xi_1 \leq \tilde{d}(T) \leq \tilde{d}(S\cup\{x\}) \leq \xi_2$ and $0 \leq b \leq 1$, we have $b(\xi_1)^{b-1} \geq b(\xi_2)^{b-1} \geq 0$. Therefore,
$$
b(\xi_1)^{b-1} \cdot (\tilde{d}(T) - \tilde{d}(T\cup\{x\})) \geq b(\xi_2)^{b-1} \cdot (\tilde{d}(S) - \tilde{d}(S\cup\{x\})),
$$
or
\begin{equation}
\label{formula:submodularity}
\begin{split}
\tilde{d}(T)^b - \tilde{d}(T\cup\{x\})^b \geq \tilde{d}(S)^b - \tilde{d}(S\cup\{x\})^b.
\end{split}
\end{equation}
Eqn. (\ref{formula:submodularity}) proves the submodularity of $\tilde{d}(S)^b$.

\underline{\textit{Case 2.}} If $\tilde{d}(T) > \tilde{d}(S\cup\{x\})$, similar to Case 1, we can prove that 
$$
\tilde{d}(S\cup\{x\})^b  - \tilde{d}(T\cup\{x\})^b \geq \tilde{d}(S)^b - \tilde{d}(T)^b.
$$
which is equivalent to Eqn. (\ref{formula:submodularity}).
\end{proof}

\noindent \textbf{Remark.} In general, the minimum or the product of two submodular functions may not be submodular. For example, let $f_1(S) = f_2(S) = |S|$. Both $f_1$ and $f_2$ are submodular, but $f_1(S)\cdot f_2(S) = |S|^2$ is not.

\end{spacing}
\end{document}